\documentclass[10pt,twocolumn,twoside]{IEEEtran}
\usepackage{graphicx}
\usepackage{latexsym}
\usepackage{epsfig}
\usepackage{pstricks}
\usepackage{amsmath,amssymb}
\usepackage{pstricks,pst-node,pst-text,pst-3d}
\usepackage[]{mcode}
\usepackage[hyphens]{url}
\usepackage{textcomp}
\usepackage{color}
\renewcommand{\baselinestretch}{1}
\begin{document}
\title{A Smartphone-Based Acquisition System for Hips Rotation Fluency Assessment}
\author{Andrea Civita$^a$, Simone Fiori$^b$, Giuseppe Romani$^a$
\thanks{
$^a$School of Information and Automation Engineering, Facolt\`{a} di Ingegneria, Universit\`{a} Politecnica delle Marche, Via Brecce Bianche, Ancona I-60131,
Italy.\newline
\indent$^b$Dipartimento di Ingegneria dell'Informazione, Facolt\`{a} di Ingegneria, Universit\`{a} Politecnica delle Marche, Via Brecce Bianche, Ancona I-60131,
Italy. (eMail: s.fiori@univpm.it). Corresponding author.\newline
\indent The present technical report is dated \today. Personal use of this material is permitted. However, permission to use this material for any other purposes must be obtained from the corresponding author.}}
\markboth{Technical report -- Dipartimento di Ingegneria dell'Informazione, Universit\`{a} Politecnica delle Marche}{Civita, Fiori, Romani: A Smartphone-Based Acquisition Systems for Hips Rotation Fluency Assessment}%
\maketitle
\def\bbbr{{{\mathbb{R}}}}
\def\bbbe{{{\mathbb{E}}}}
\def\bbbu{{{\mathbb{N}}}}
\def\bbbg{{{\mathbb{G}}}}
\def\D{{{\mathrm{D}}}}
\def\d{{\mathrm{d}}}
\def\mdef{\,{\stackrel{{\mathrm{def}}}{=}}\,}
\def\tr{\mathrm{tr}}
\def\mass{{{m}}}
\def\P{{{\mathrm{P}}}}
\def\R{{{\mathrm{R}}}}
\def\bbbi{{{\mathbb{I}}}}
\def\one{{{\mathbf{1}}}}
\def\Exp{{{\mathrm{Exp}}}}
\def\Log{{{\mathrm{Log}}}}
\def\MATLAB{\textsc{Matlab}$^\circledR$}
\def\Android{{\textsc{Android}\textsuperscript{TM}}}
\def\Windows{{\textsc{Microsoft$^\circledR$ Windows$^\circledR$}}}
\begin{abstract}
The present contribution is motivated by recent studies on the assessment of the fluency of body movements during complex motor tasks. In particular, we focus on the estimation of the Cartesian kinematic jerk (namely, the derivative of the acceleration) of the hips' orientation during a full three-dimensional movement. The kinematic jerk index is estimated on the basis of gyroscopic signals acquired through a smartphone. A specific free mobile application available for the \Android\ mobile operating system, Hyper IMU, is used to acquire the gyroscopic signals and to transmit them to a personal computer via a User Datagram Protocol (UDP) through a wireless network. The personal computer elaborates the acquired data through a \MATLAB\ script, either in real time or offline, and returns the kinematic jerk index associated to a motor task.
\end{abstract}
\section{Introduction}\label{sec1}
\begin{figure}[h]
\centering
\includegraphics[width=0.7\columnwidth, height=0.9\columnwidth]{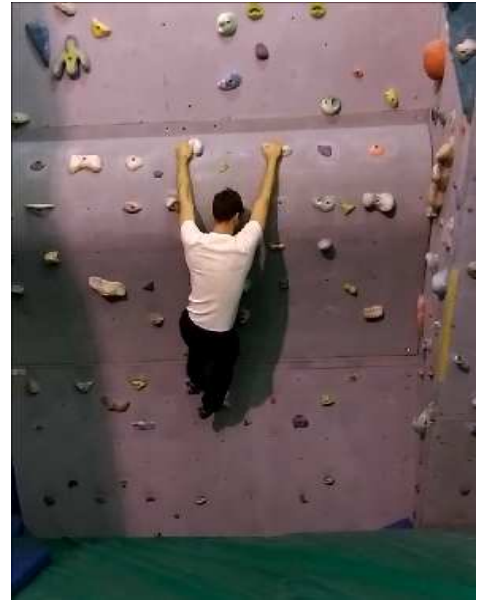}
\caption{Illustration of an indoor wall-climbing setup.}  
\label{fig:climbing}
\end{figure}

\PARstart{C}{omplex} body tasks, such as rock climbing, involve the alternation of periods dedicated to postural regulation and of quadruped displacement. The Figure~\ref{fig:climbing} illustrates a typical hips sway during climbing. In neurosciences engineering \cite{biess} and in sports science \cite{russell}, the assessment of body fluency provides valuable information on the patient's musculoskeletal system as well of the mental state of an athlete (for example, rigid posture and erratic movements during a climbing performance may indicate a state of anxiety \cite{seifert}); likewise, since restoring movement fluency is a key focus in physical rehabilitation, objective measurement of fluency is of prime importance \cite{kerr}. 

While previous studies on the fluency of body movements from temporal and spatial measurement analyses took into account only the displacement of the hips on a two-dimensional projective plane \cite{cordier}, recent studies have highlighted how the presence of anteroposterior and lateral sway would rather imply the need of combined temporal and full 3-dimensional movement analysis. Such analysis should take into account both hip translation and rotation. The recent study \cite{seifert}, in particular, focuses on the assessment of the \emph{kinematic jerk} of the hips rotation during a complex body task. In kinematics, the jerk of a positional variable is defined as its third time-derivative, namely, as the derivative of the acceleration. 

The hips orientation is described by a $3\times 3$ rotation matrix which is a function of the time. Therefore, the orientation jerk is defined as the third-order time-derivative of a curve on the group of special orthogonal matrices. Such group, denoted as $\mathrm{SO}(3)$, is a curved manifold, hence, in such space, the assessment of kinematic variables such as acceleration and jerk is not straightforward as it is for curves in the ordinary space $\bbbr^3$. 

Although a complete analysis of body movement fluency would imply a statistical characterization of both hips rotation jerk and translation jerk, the study \cite{seifert} showed that the (normalized) kinematic jerk of hips orientation and the (normalized) kinematic jerk of hips trajectory exhibit a significant positive correlation, signifying that both measures provide a similar measure of fluency. In addition, several clinical studies focus on hips rotation only, such as, for example, the research \cite{kettunen}, that focuses on the analysis of reduced range of motion of the hips, which is an established clinical indicator of osteoarthritis.

A number of related studies made use of the kinematic jerk of positional data to investigate (bio)mechanical phenomena. 
In the studies \cite{biess,flash}, concerning human three-dimensional pointing movements, it is assumed that the speed profile along the hand path is determined by minimization of the squared kinematic jerk. Likewise, in the paper \cite{rubio}, the impact of a kinematic jerk index is analyzed in the context of the the generation of minimum time collision-free trajectories for industrial robots in a complex environment. 
In the research paper \cite{tack}, it was hypothesize that there exists a functional relationship between kinematic jerk cost and energy consumption during walking. Energy consumption during walking is closely related to the gait speed. Minimum kinematic jerk theory states that a body movement is carried out by minimizing the kinematic jerk of the motion trajectory.
The study \cite{he} showed that kinematic jerk and its response spectrum can enhance the recognition of the nonstationary groundmotion. In fact, kinematic jerk represents the nonstationary component in the high frequency band of an earthquake wave.

The aim of the present contribution is to illustrate the details of a smartphone-based acquisition systems to evaluate hips rotation fluency by estimating the kinematic jerk of the acquired signals. Nowadays, smartphones come with a large set of embedded sensors, including gyroscopes. The previous contribution \cite{mellone} investigated instrumented `Timed Up and Go' based on a specialized measurement system, namely, the accelerometer of a smartphone, to identify and evaluate specific mobility skills. The contribution \cite{nishiguchi} evaluated the reliability of a smartphone accelerometer in gait analysis. Likewise, the project SmartGait$^\textsuperscript{TM}$ \cite{purdue} makes use of a smartphone to acquire a number of postural indicators from a moving body with the aim to develop an inexpensive and accurate gait tracking tool that can be easily implemented by researchers and clinicians.
\section{Acquisition of a gyroscopic data-stream by a smarthphone and a personal computer}
The present section explains how to connect wirelessly an {\Android}-based smartphone with a personal computer equipped with a \MATLAB\ development environment to  process the acquired data.

\begin{figure}[h]
\centering
\includegraphics[width=0.9\columnwidth, height=0.6\columnwidth]{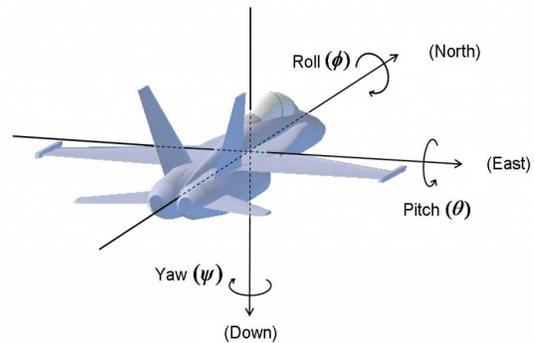}
\caption{An illustration of the three angles yaw, pitch and roll returned by the `Hyper IMU' mobile application.} 
\label{Yaw-Pitch-Roll}
\end{figure}

\begin{figure}[h]
\centering
\includegraphics[width=0.9\columnwidth, height=1.4\columnwidth]{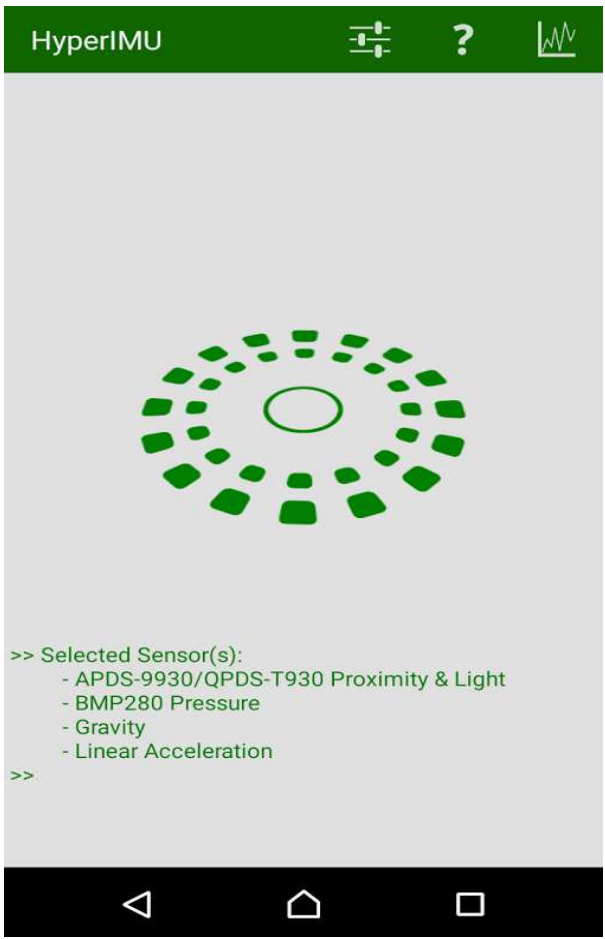}
\caption{A screenshot of the freeware `Hyper IMU' mobile application for the \Android\ mobile operating system.} 
\label{hyperimu}
\end{figure}

The smartphone runs under an \Android\ mobile operating system and is worn on the waist of an athlete or a patient. An installed mobile application, namely the Hyper IMU, version 1.3, acquires the orientation of the smartphone expressed by three angles, namely \emph{yaw}, ranging in $[0,\ 360)$, \emph{pitch}, ranging in $[-180,\ 180)$ and \emph{roll}, ranging in $[-90,\ 90]$. An illustration of the three angles is displayed in the Figure~\ref{Yaw-Pitch-Roll}. 
%
The application allows the user to set the sampling period from a minimum value of $\Delta T = 20$ms. A screenshot of the Hyper IMU mobile application is displayed in the Figure~\ref{hyperimu}. 

The transmission of the data-stream from a smartphone to a {\MATLAB}-equipped personal computer takes place through a User Datagram Protocol (UDP, RFC 768) \cite{postel}. The UDP was defined to make available a simple packet-switched communication mode in networks of interconnected devices. The UDP assumes that the Internet  Protocol  (IP) is used as underlying protocol. The UDP provides a means for mobile applications and computer programs to exchange messages to each other  with a minimum of protocol machinery. 

If the room where the measures take place is served by a wireless router, the smartphone and the personal computer may be connected to the wireless network and exchange data. Otherwise, (as, for example, in an outdoor experiment), it is possible to make a personal computer behave as a wireless-network access point that the smartphone may bind to. Under a \Windows\ operating system, the access point may be created upon the execution of the command
\begin{verbatim}
netsh wlan set hostednewtork mode allow 
ssid=... key=... keyUsage=persistent
\end{verbatim}
where the parameter \texttt{ssid} denotes the name assigned to the access point and the parameter \texttt{key} denotes the access password. The wireless network starts operating upon the execution of the command
\begin{verbatim}
netsh wlan start hostednetwork
\end{verbatim}
while, upon the execution of the command
\begin{verbatim}
netsh wlan stop hostednetwork
\end{verbatim}
the access point stops operating.

The \MATLAB\ platform was used to acquire the gyroscopic signals from the smartphone and to compute the kinematic jerk index of an acquired signal. In the \MATLAB\ language, a port is managed like a data-file, therefore, reading out signals from a smartphone via a wireless network can be performed like reading the records of a data-file, as shown in the code excerpt Listing~\ref{lst:acquisition}.
\begin{lstlisting}[numbers=right,firstnumber=1,float=h,label=lst:acquisition,caption=\MATLAB\ script to acquire the gyroscopic signals from the smartphone via a wireless network.\newline]
% MATLAB code to read data from a WiFi network by 
% the UDP protocol by Andrea Civita, Giuseppe 
% Romani and Simone Fiori (Universita' Politecnica 
% delle Marche, October 2015)
DataStream = 99999; LstTime = 180; Lport = 1234; 
UDPComIn = udp('0.0.0.0','LocalPort',LPort); 
set(UDPComIn,'DatagramTerminateMode','off',...
    'InputBufferSize',DataStream,'TimeOut',...
    LstTime) 
fopen(UDPComIn); 
readout = fscanf(UDPComIn,'%s',DataStream); 
fclose(UDPComIn);
scandata = textscan(readout,'%s',DataStream,...
                    'Delimiter',', ,#');
scan = scandata{1};
scan=scan(~cellfun('isempty',scan));
apparray = cellfun(@str2num,scan);
data = vec2mat(apparray,3)';

\end{lstlisting}
Such script shows the minimal instructions set necessary to acquire the data correctly. It might need additional instructions to reset the \MATLAB\ settings (such as the \texttt{instrreset} command) and to display and save the acquired data.

Line 6 of the \MATLAB\ script creates the UDP object \texttt{UDPComIn} and specifies the IP address and the port number \texttt{Lport} of the transmitter. Setting the IP address to the conventional value \texttt{0.0.0.0} indicates that the actual IP address of the transmitting device is not specified and that the computer will listen to any data stream sent through the port specified at Line 5 (which needs, therefore, to be uniquely dedicated to the communication between the smartphone and the personal computer). Using the conventional IP address \texttt{0.0.0.0} allows the acquisition script in Listing~\ref{lst:acquisition} to work within a dynamic host configuration protocol (DHCP) network.

The Line 7 of the script sets relevant attributes of the object \texttt{UDPComIn}, namely:
\begin{itemize}
\item the attribute \texttt{UDPComIn.DatagramTerminateMode} to the value \texttt{off}, in order to make the personal computer keep acquiring packets as they become available from the smartphone;
\item the attribute \texttt{UDPComIn.InputBufferSize} to the value \texttt{DataStream}, whose value is set in Line 5, so that the personal computer will stop listening after a sufficiently large number of acquired characters;
\item the attribute \texttt{UDPComIn.TimeOut} to the value \texttt{LstTime}, expressed in seconds and set in Line 5, so that the personal computer will stop listening after a sufficiently time-span to allow acquiring the gyroscopic signals pertaining to a sufficiently long experiment.
\end{itemize}
It should be noted that the conditions set on the input buffer size and on the timeout cause the listening process to stop, whichever verifies first.

Line 11 of the \MATLAB\ script Listing~\ref{lst:acquisition} reads the character-string output from the smartphone into the variable \texttt{readout}. The variable \texttt{readout} appears as a string of the format\\\\
Yaw$_1$,Pitch$_1$,Roll$_1$\#Yaw$_2$,Pitch$_2$,Roll$_2$\#Yaw$_3$,Pitch$_3$,Roll$_3$...\\\\
Namely, every record is formed by three angles (expressed in degrees and represented as floats) separated by two commas and each record ends with a hash character.

The commands in the Lines 13-15 separate each record and construct a cell object, named \texttt{scan}, that contains the list of acquired yaw, pitch and roll angles. Line 16 expunges empty records.

The Line 17 converts the string-type cell elements into numbers and the Line 18 constructs a $3\times N$ array \texttt{data} whose columns contain the acquired angles.

A further auxiliary \MATLAB\ script converts the yaw, pitch and roll angles triples to a gyroscopic signal $R(t)$, whose elements are $3\times 3$ rotation matrices, that represents the instantaneous orientation of the smartphone with respect to a fixed reference frame (typically defined on the basis of a magnetoscopic sensor). The obtained gyroscopic signal will be converted back to angular speeds as explained in the next Section. The intermediate rotation data are introduced in order to make the procedure independent of the type of acquired data, since different gyroscopic mobile applications might acquire different kinds of angles\footnote{The \emph{proper Euler angles triples} and the \emph{Tait-Bryan angles triples} amount to twelve possible combinations \cite{Greenwood}.}.
\section{Computational estimation of the kinematic jerk index associated to a gyroscopic data stream}\label{sec2}
%
Given a hips rotation trajectory $R(t)$, define the ``angular velocity'' matrix field
\begin{equation}
\Omega(t)=R^T(t)\frac{dR(t)}{dt}.
\end{equation} 
The matrix $\Omega$ is $3\times 3$ and skew-symmetric, namely, its diagonal elements are zero and only its $3$ strictly-upper triangular entries are independent of each other, namely
\begin{equation}
\Omega(t)=\left[
           \begin{array}{ccc}
            0 & \omega_a(t) & \omega_b(t)\\
            -\omega_a(t) & 0 & \omega_c(t)\\
            -\omega_b(t) & -\omega_c(t) & 0
           \end{array}
          \right].
\end{equation}
The ``vee'' operator $^\vee$ \cite{greg} returns, in a vector, such independent entries. One may, thus, associate to a trajectory $R(t)$ a vectorized angular velocity 
\begin{equation}
 \omega(t)\mdef \Omega^\vee(t)=[R^T(t)\dot{R}(t)]^\vee,
\end{equation} 
whose entries are $\omega(t)=[\omega_a(t)\ \omega_b(t)\ \omega_c(t)]^T$.

The contribution \cite{seifert} takes, as hips rotation jerk, the second-order time-derivative of the $3\times 1$ vector $\omega$, namely $\ddot{\omega}(t)$. The advantage of such definition of kinematic jerk is that it appears as a vector in $\bbbr^3$ and the scalar kinematic jerk may be simply computed through the 2-norm $\|\ddot{\omega}\|$. By using the above parameterization for the angular velocity matrix, the scalar kinematic jerk taken in \cite{seifert} is computed as
\begin{equation}\label{ssjerk}
\|\ddot{\omega}\|=\sqrt{\ddot{\omega}_a^2+\ddot{\omega}_b^2+\ddot{\omega}_c^2}. 
\end{equation}

The integrated kinematic jerk index associated to a trajectory $R(t)$, $t\in[0\ T]$, is defined as
\begin{equation}\label{Jdef}
J\mdef C\int_{0}^T\|\ddot{\omega}\|dt,
\end{equation}
where the constant $C$ is chosen in a way that makes the kinematic jerk index $J$ dimensionless. The index $J$ defined in (\ref{Jdef}) is termed \emph{Cartesian kinematic jerk} to distinguish it from other possible definitions (see, for example, \cite{Tsao}). Furthermore, the constant $C$ is chosen in a way that makes the kinematic jerk index  independent from the total length of a path and from the total observation time $T$.

In order to estimate numerically the derivative $\dot{R}(t)$, the paper \cite{seifert} makes implicitly use of the following idea. Let $R_k$ and $R_{k+1}$ denote two subsequent samples of an acquired, discrete-time gyroscopic data-stream counting $N$ observations spaced apart in time of $\Delta T$. The relationship between the number of acquired samples, the sampling interval and the total observation time is 
\begin{equation}\label{time}
 T=(N-1)\Delta T. 
\end{equation} 
Consider such samples as joined by a geodesic arc of the curved space $\mathrm{SO}(3)$ (for a review of the geometry of the special orthogonal group, readers might consult, e.g., \cite{eas,ft:2009}), namely 
\begin{equation}
 R_{k+1}=\exp_{R_{k}}(\Delta T\cdot R_k\cdot\Omega_{k})=R_k\Exp(\Delta T\cdot \Omega_k),
\end{equation} 
where $\exp_R(\cdot)$ denotes a group exponential map at $R\in\mathrm{SO}(3)$ and $\Exp$ denotes the matrix exponential.
Reversing such formula by means of the logarithmic map, one gets the estimate
\begin{equation} 
 \Omega_k=\frac{1}{\Delta T}\Log(R_k^TR_{k+1}),\ k=0,\ 2,\ \ldots,\ N-1,
\end{equation}
where $\Log$ denotes the (principal) matrix logarithm (that can be calculated, by the \MATLAB\ language, by the function \texttt{logm}).
Let us define $\omega_k\mdef\Omega_k^\vee$. Now, the kinematic jerk index may be estimated as the sum
\begin{equation}\label{Jcalc}
\tilde{J}\mdef C\Delta T\sum_{k=2}^{N-1}\|\ddot{\omega}_k\|,
\end{equation}
where 
\begin{equation}
\ddot{\omega}_k\mdef \frac{\omega_k-2\omega_{k-1}+\omega_{k-2}}{(\Delta T)^2}.
\end{equation}
The total length of a measured data-stream may be defined as
\begin{equation}
d\ \mdef \int_{0}^T\|\omega\|dt,
\end{equation}
which can be estimated numerically as
\begin{equation}\label{dist}
\tilde{d}\ \mdef \Delta T\sum_{k=0}^{N-1}\|\omega_k\|.
\end{equation}
In the spirit of the contributions \cite{seifert,Tsao}, the normalization constant $C$ may be defined as
\begin{equation}
C\mdef \frac{(N-2)^2\Delta T^2}{\tilde{d}}.
\end{equation}

The \MATLAB\ function that implements the above calculations is displayed in the Listing~\ref{lst:jerk}.
\begin{lstlisting}[numbers=right,firstnumber=1,float=h,label=lst:jerk,caption=\textsc{Matlab}$^\circledR$ function to calculate the Cartesian kinematic jerk.]
% Function to compute the Cartesian jerk by 
% Andrea Civita, Giuseppe Romani and Simone 
% Fiori (Universita' Politecnica delle Marche, 
% October 2015)
function J = jerk(R,DT)
L = size(R,3);
z = zeros(3,L-1); 
for k=2:L,
 Omega = (1/DT)*real(logm(R(:,:,k-1)'*R(:,:,k)));
 z(:,k-1) = [Omega(1,2),Omega(1,3),Omega(2,3)]; 
end
d = T*sum(sqrt(sum(z.^2,1)));
je = (1/DT^2)*diff(diff(z,1,2));
Cj = (L-3)^2*DT^2/d;
J = Cj*DT*sum(sqrt(sum(je.^2,1)));


\end{lstlisting}

The \MATLAB\ function displayed in the Listing~\ref{lst:jerk} inputs an array \texttt{R} of size $3\times3\times N$ and a sampling period \texttt{DT} and returns the numerical estimate of the Cartesian kinematic jerk index (\ref{Jcalc}). At Line 9, note that the operator \texttt{real} was introduced to get rid of the imaginary part of the matrix logarithm, that, in principle, should be equal to zero, while, in practice, it arises from finite-precision machine calculations.
\section{Experimental results}\label{sec4}
The current Section presents results of experiments on the acquisition of hips rotation data and on the computational assessment of the fluency of a body movement during complex motor tasks. The experiments were performed on acquired data measured in different conditions.
%
\subsection{Acquisition of test-type complex-body-movement signals}\label{ss43}
As a preliminary test of the described acquisition procedure and of the kinematic jerk-estimation scripts, three different kinds of gyroscopic signals were acquired during three complex body tasks: walking, running and jumping. The walking and running body tasks were carried out over a straight path $10$ meters long. During the acquisition of the gyroscopic signals, the smartphone was fastened to the subject's belt. 

\begin{figure}[h]
\centering
\includegraphics[width=1\columnwidth, height=0.8\columnwidth]{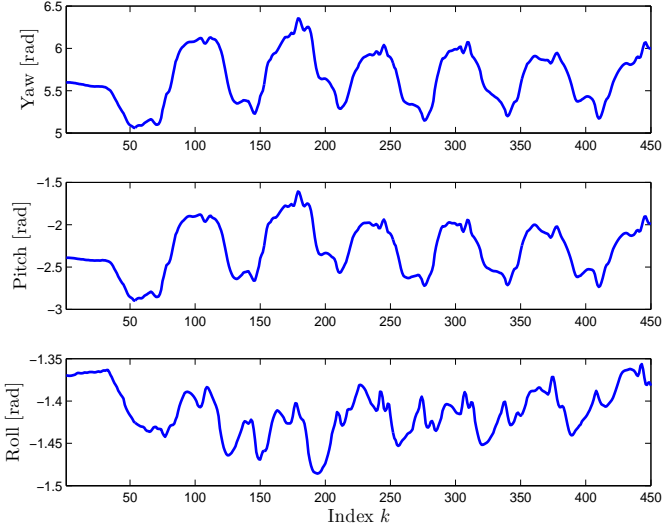}
\caption{Walking, running and jumping gyroscopic signal. Graphical representation of the Yaw, Pitch and Roll angles acquired during \emph{walking} (unwrapped, radians).} 
\label{Walk}
\end{figure}

\begin{figure}[h]
\centering
\includegraphics[width=1\columnwidth, height=0.8\columnwidth]{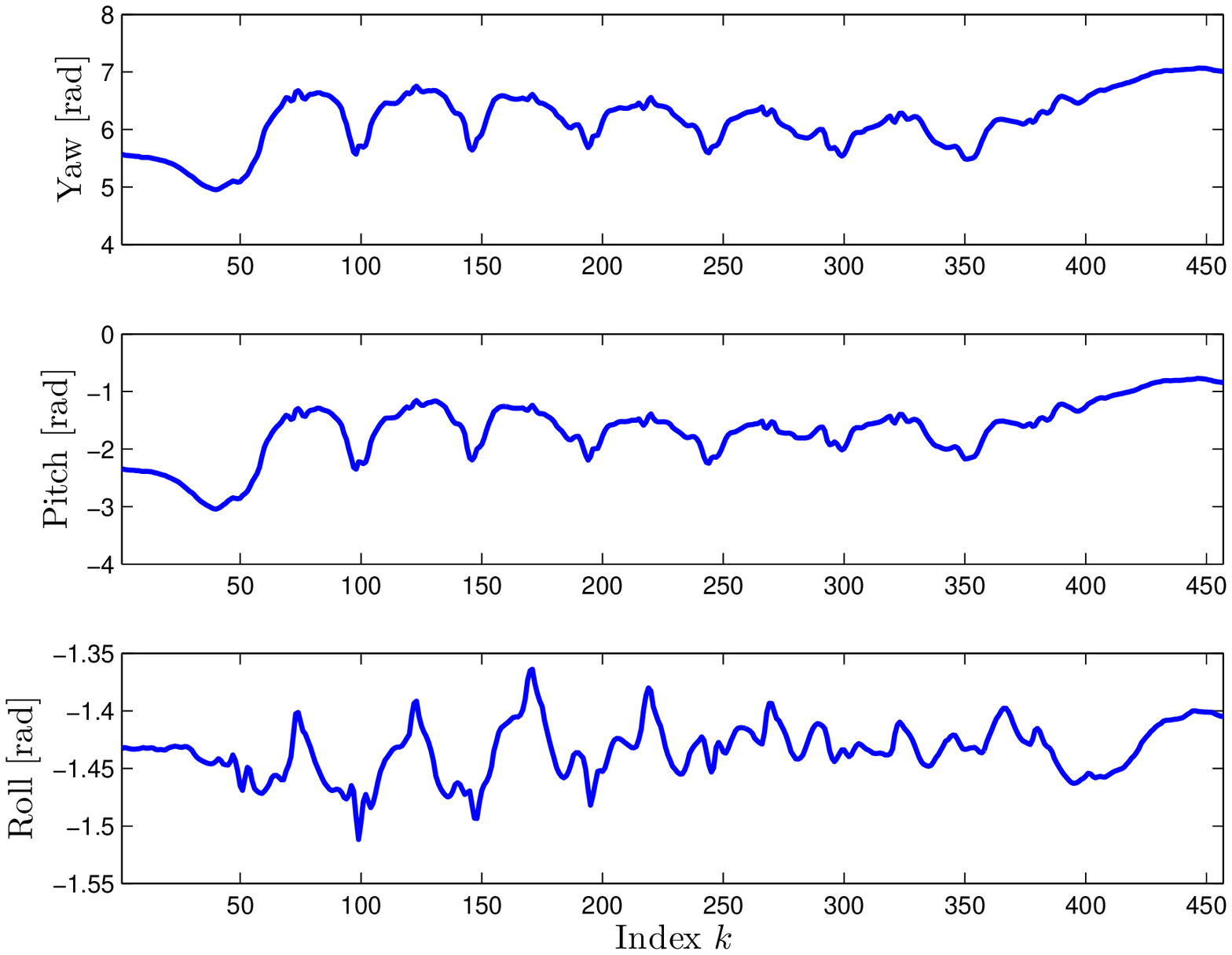}
\caption{Walking, running and jumping gyroscopic signal. Graphical representation of the Yaw, Pitch and Roll angles acquired during \emph{running} (unwrapped, radians).} 
\label{Run}
\end{figure}

\begin{figure}[h]
\centering
\includegraphics[width=1\columnwidth, height=0.8\columnwidth]{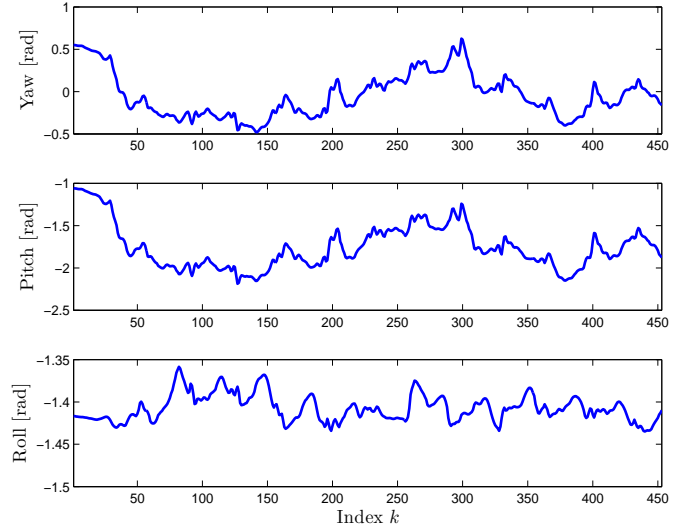}
\caption{Walking, running and jumping gyroscopic signal. Graphical representation of the Yaw, Pitch and Roll angles acquired during \emph{jumping} (unwrapped, radians).} 
\label{Jump}
\end{figure}

A number $N=450$ of samples was acquired for each task. The data stream acquired during walking is displayed in the Figure~\ref{Walk}, the data stream acquired during running is displayed in the Figure~\ref{Run} and the data stream acquired during jumping is displayed in the Figure~\ref{Jump}.

The results of the computation of the Cartesian kinematic jerk $\tilde{J}$ according to the formula (\ref{Jcalc}) are displayed in the Table~\ref{Experiment2-table}. As it is readily appreciated, the Cartesian kinematic jerk grows in value as the  analyzed motion becomes increasingly erratic.
\begin{table}[h]
\centering
\begin{tabular}{|l|c|}
\hline
&\textbf{Cartesian kinematic jerk}\\\hline
\textbf{Walking}&104517\\\hline
\textbf{Running}&158509\\\hline
\textbf{Jumping}&206153\\\hline
\end{tabular}
\caption{Cartesian kinematic jerk calculated on acquired gyroscopic signals of the type `Walk', `Run' and `Jump'.}
\label{Experiment2-table}
\end{table}
\subsection{Acquisition of signals during indoor wall climbing sessions}\label{ss44}
The gyroscopic signals pertaining to the present experiment were acquired in a gym equipped with a professional wall-climbing facility and the signals were acquired by the help of three athletes of different experience. 

In particular, an Expert, an Intermediate-level and an Inexpert climbers completed two different climbing tasks, hereafter termed \emph{Task $\alpha$} and \emph{Task $\beta$}. A frame of a camera-recording taken during a climbing session is displayed in the Figure~\ref{fig:climbing}. The experience level of the athletes as well as the hardness of a climbing task were evaluated by a professional trainer. 

Before performing each task, the athletes planned a sequence of hand-grasps, in order to make the execution of the tasks as uniform as possible, while the foot standing strategy was left to each athlete's discretion. Nevertheless, due to the different level of experience of the athletes, the elapsed time took by each athlete to complete the same task was different, resulting in a different data-stream length for each of the six resulting sessions (one session per athlete and per task).

\begin{figure}[h]
\centering
\includegraphics[width=1\columnwidth, height=0.8\columnwidth]{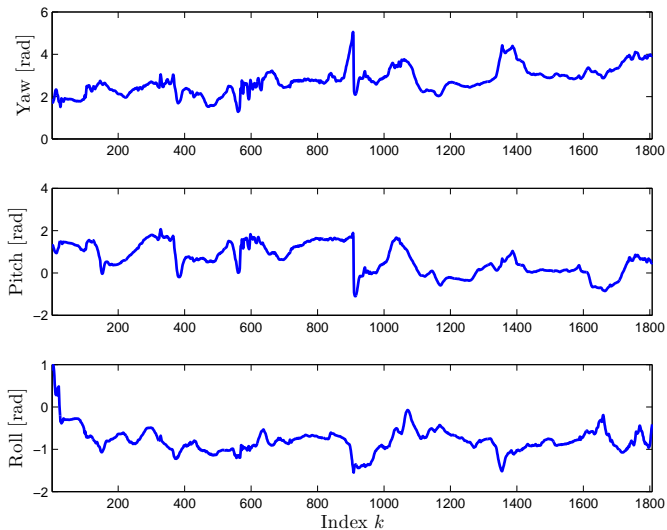}
\caption{Gyroscopic signals acquired in a gym equipped with a professional wall-climbing facility: Inexpert athlete on Task $\beta$ (unwrapped, radians).} 
\label{Inexpert}
\end{figure}

\begin{figure}[h]
\centering
\includegraphics[width=1\columnwidth, height=0.8\columnwidth]{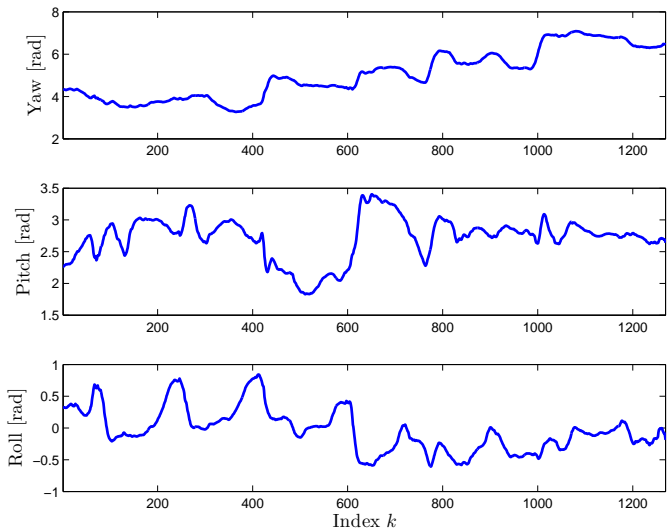}
\caption{Gyroscopic signals acquired in a gym equipped with a professional wall-climbing facility: Intermediate-level athlete on Task $\beta$ (unwrapped, radians).} 
\label{Medium}
\end{figure}

\begin{figure}[h]
\centering
\includegraphics[width=1\columnwidth, height=0.8\columnwidth]{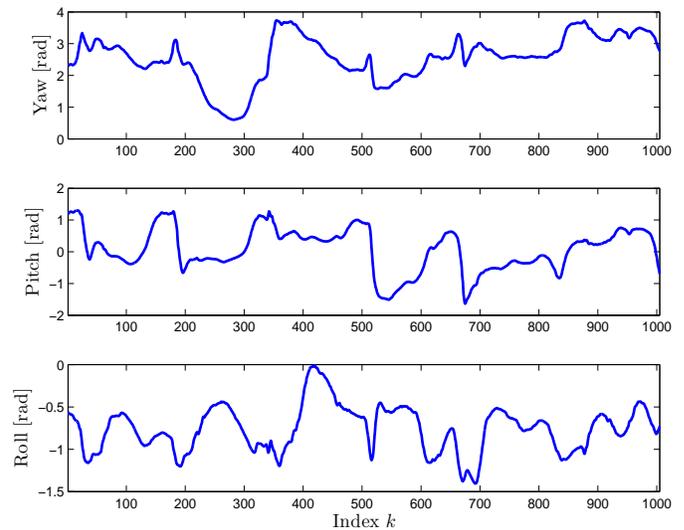}
\caption{Gyroscopic signals acquired in a gym equipped with a professional wall-climbing facility: Expert athlete on Task $\beta$ (unwrapped, radians).} 
\label{Expert}
\end{figure}

As an example of the acquired signals over Task $\beta$, the acquired gyroscopic signals, under the form of Yaw, Pitch and Roll pertaining to the Inexpert athlete are displayed in the Figure~\ref{Inexpert}, the acquired signals pertaining to the Intermediate-level athlete are displayed in the Figure~\ref{Medium}, while the gyroscopic signals pertaining to the Expert athlete are displayed in the Figure~\ref{Expert}. 

The results of Cartesian kinematic jerk computation are displayed in the Table~\ref{Experiment3-table}. Even in this case, for each task separately, the Cartesian kinematic jerk grows in value as the analyzed motion becomes less fluent. It is interesting to compare the value of the accumulated kinematic jerk $\tilde{J}$ index with the value of the accumulated velocity $\tilde{d}$ also reported in the Table~\ref{Experiment3-table}: 
Such phenomenon may be rephrased in terms of kinematics statistics by saying that there exists a weak correlation between the velocity index and the kinematic jerk index, in general.
\begin{table}[h]
\centering
\begin{tabular}{|l|c|c|}
\hline
&\textbf{Distance (\ref{dist})}&\textbf{Kinematic jerk (\ref{Jcalc})}\\\hline
\textbf{Inexpert -- Task $\alpha$}&78&3190177\\\hline
\textbf{Intermediate-level -- Task $\alpha$}&28&1136228\\\hline
\textbf{Expert -- Task $\alpha$}&41&695616\\\hline
\textbf{Inexpert -- Task $\beta$}&39&797533\\\hline
\textbf{Intermediate-level -- Task $\beta$}&24&674140\\\hline
\textbf{Expert -- Task $\beta$}&22&377639\\\hline
\end{tabular}
\caption{Distance (\ref{dist}) and Cartesian kinematic jerk (\ref{Jcalc}) calculated on acquired gyroscopic signals for athletes `Inexpert', `Intermediate-level' and `Expert' on both `Task $\alpha$' and `Task $\beta$'.}
\label{Experiment3-table}
\end{table}
\section{Conclusion}\label{sec5}
The present research work deals with a new acquisition procedure to evaluate hips orientation fluency during a complex body motion task. The illustrated data-acquisition procedure is based on a smartphone endowed with a gyroscopic sensor that is able to reveal the instantaneous orientation of the hips of an athlete during motion. The smartphone runs an application that sends out the measured data over a wireless network and that transmits the acquired data, according to a User Datagram Protocol, to a personal computer. 

The personal computer, through a computer program written in \MATLAB\ language and interpreted by a \MATLAB\ environment, acquires in real time the gyroscopic data-stream and evaluates numerically a kinematic jerk index, taken as a meaningful measure of body movement fluency. The illustrated procedure was first tested on specific motor tasks, such as walking, running and jumping, as a validation stage. Then, data acquisitions were performed in a professional gymnasium, equipped with a professional wall-climbing facility. In particular, the hips orientation during climbing of three athletes with different skills were recorded over two different walls. 

The obtained results confirm that the procedure is reliable and the numerical results of kinematic jerk index estimation could confirm the self-evaluation of the athletes' skills.

Possible extensions of the present work are real-time processing of the acquired gyroscopic data (to allow athletes to obtain a real-time feedback on their performances), multichannel acquisition by several concurring smartphones (to allow merging gyroscopic data pertaining to different body movements), remote acquisition/processing of gyroscopic data (to allow a central server to acquire and process the data streams sent over a wireless or cable network from different remote geographical sites) in a telemedicine setting, and the design of an automated (possible machine-learning based) classification algorithm to discern between different body-movement fluency levels.
\section*{Acknowledgment}
The authors wish to thank Ludovic Seifert for sharing his papers on motion fluency estimation, Edoardo Russo and Nicola Sabino, from the School of Information and Automation Engineering, Universit\`{a} Politecnica delle Marche, and Hwee Kuan Lee from the Bio-Informatics Istitute (A*STAR, Republic of Singapore), for constructive discussions during the development of the present project. 
\vspace{-1cm}
\begin{biography}[{\includegraphics[width=1.1in,height=1.3in,clip]{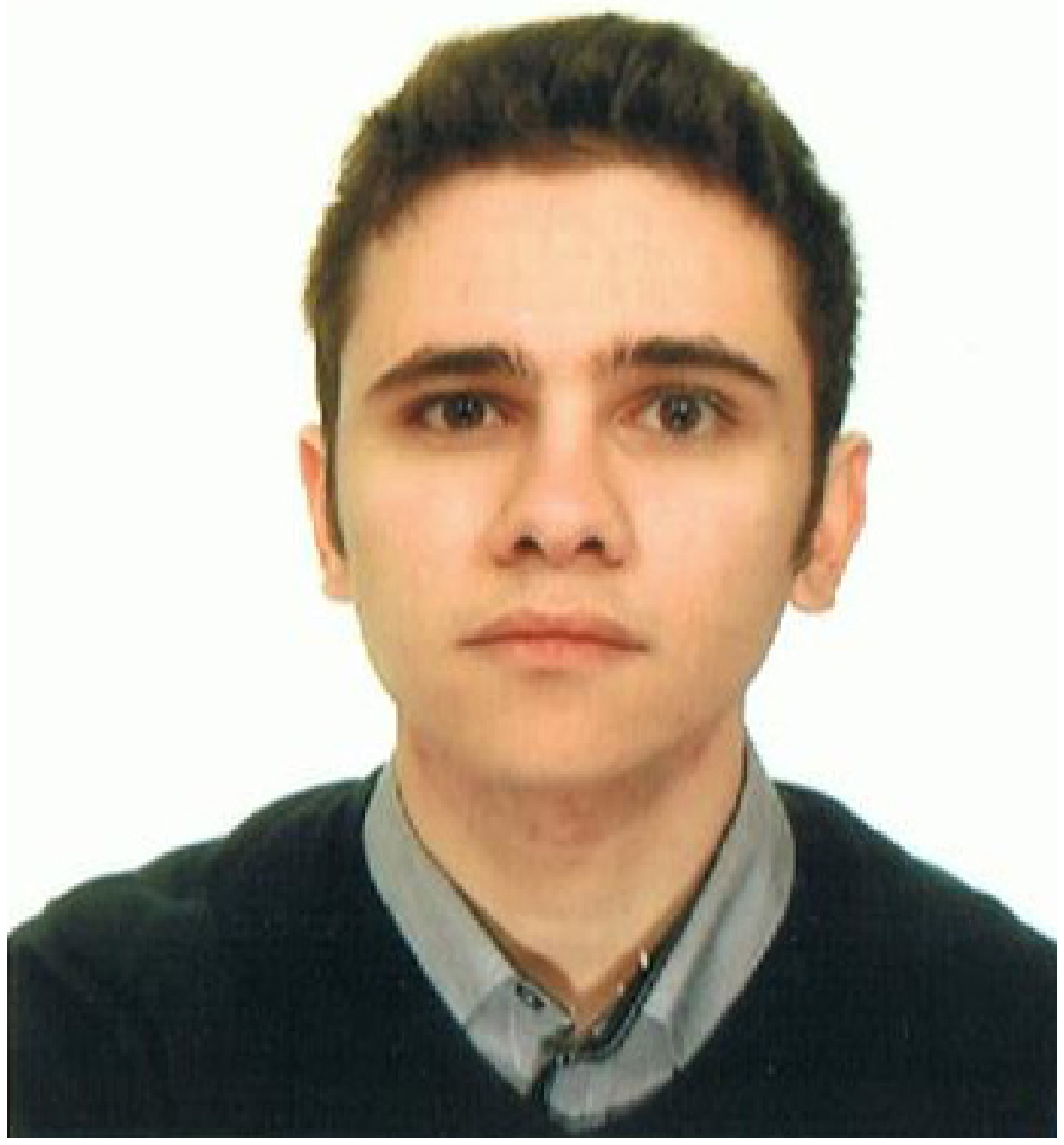}}]{Andrea Civita} is pursuing his B.Sc.\ degree at the School of Information and Automation Engineering of the Universit\`{a} Politecnica delle Marche (Ancona, Italy). 
\end{biography}
\vspace{-1.0cm}
\begin{biography}[{\includegraphics[width=1.1in,height=1.3in,clip]{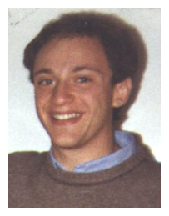}}]{Simone Fiori} received the Italian Laurea (Dr.\ Eng.)
with honors in electronics engineering in July 1996 from the University of Ancona (Italy), and the Ph.D.\ degree in electrical engineering (circuit theory) in March 2000 from the University of Bologna (Italy). In November 2005, he joined the Universit\`{a} Politecnica delle Marche. His research interests include linear and non-linear adaptive discrete-time filter theory and geometrical methods for machine learning and signal processing. He is author of about $90$ refereed journal papers and about $80$ conference/workshop papers. Dr.\ Fiori was the recipient of the 2001 ``E.R. Caianiello Award'' for the best Ph.D.\ dissertation in the artificial neural network field and the 2010 ``Rector Award'' as a proficient researcher. He is currently serving as Associate Editor of Neurocomputing (Elsevier)
and Cognitive Computation (Springer). Dr. Fiori was awarded several scholarships to visit the AIST research institute (Tsukuba, Ibaraki prefecture, Japan), the RIKEN research institute (Wako-shi campus, Saitama prefecture, Japan), the Tokyo University of Agriculture and Technology (Koganei campus, Tokyo, Japan), the Trondheim Technical University (Trondheim, Norway), the University of Bari (Bari, Italy), the Bioinformatics Institute (A*STAR - Biopolis, Republic of Singapore), the National Taiwan University (Taipei, Republic of China - Taiwan) and the Beijing Institute of Technology (Beijing, People's Republic of China). He is a member of the APSIPA.
\end{biography}
\vspace{-1cm}
\begin{biography}[{\includegraphics[width=1.1in,height=1.3in,clip]{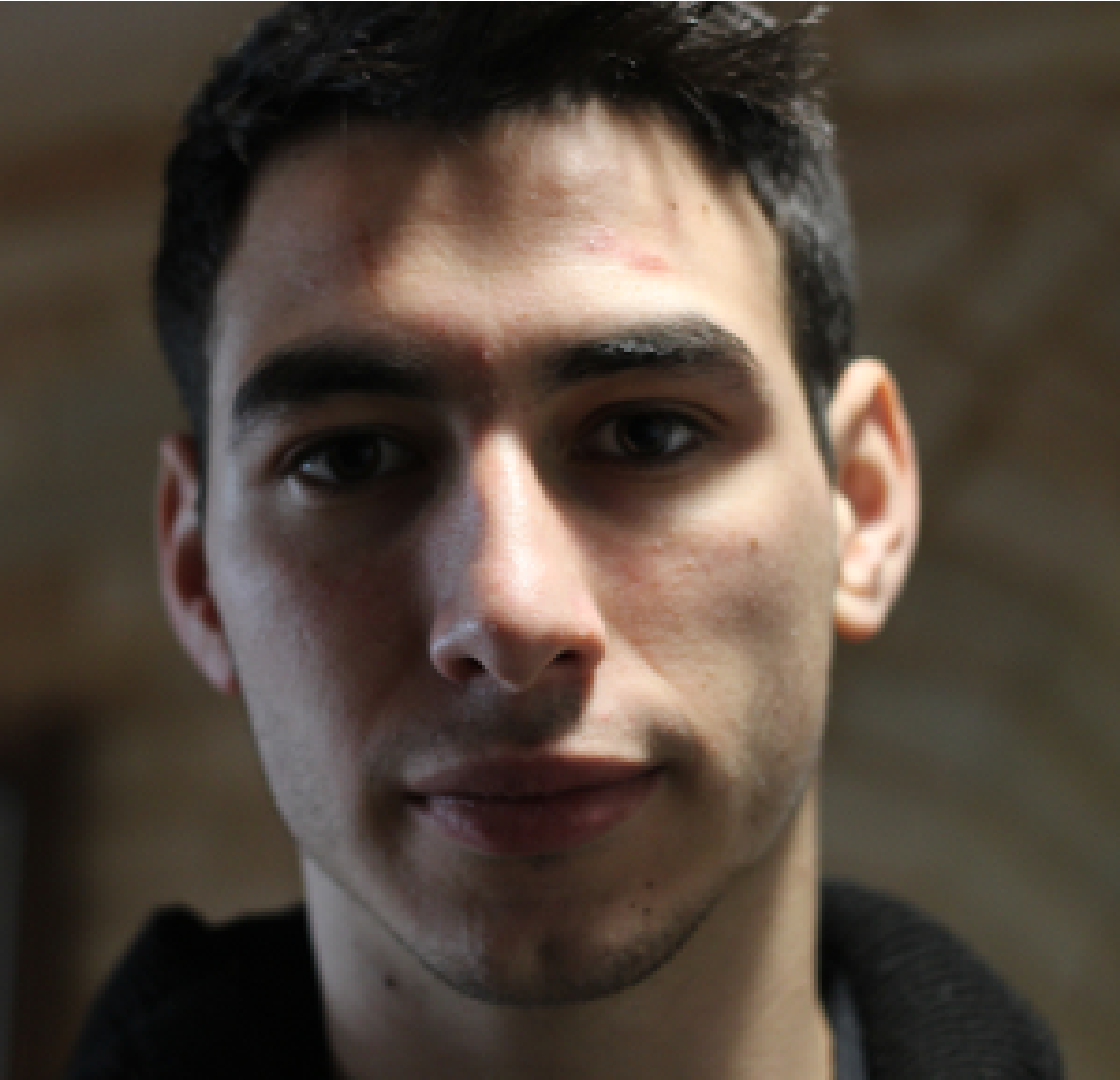}}]{Giuseppe Romani} is pursuing his B.Sc.\ degree at the School of Information and Automation Engineering of the Universit\`{a} Politecnica delle Marche (Ancona, Italy).
\end{biography}

\begin{thebibliography}{99}
%
%
\bibitem{biess} \textrm{A. Biess, D.G. Liebermann and T. Flash}, \textit{A computational model for redundant human three-dimensional pointing movements: Integration of independent spatial and temporal motor plans simplifies movement dynamics}, The Journal of Neuroscience, Vol. 27, No. 48, pp. 13045 -- 13064, November 2007
%
\bibitem{greg} \textrm{G.S. Chirikjian}, \textit{Stochastic Models, Information Theory, and Lie Groups, Volume 2: Analytic Methods and Modern Applications}, Birkh\"{a}user, 1st Edition (2011)
%
\bibitem{cordier} \textrm{P. Cordier, G. Dietrich and J. Pailhous}, \textit{Harmonic analysis of a complex motor behavior}, Human Movement Science, Vol. 15, pp.  789 -- 807, 1996
%
\bibitem{eas} \textrm{A. Edelman, T.A. Arias and S.T. Smith}, \textit{The geometry of algorithms with orthogonality constraints}, SIAM Journal on Matrix Analysis Applications, Vol. 20, No. 2, pp. 303 -- 353, 1998
%
%
%
\bibitem{ft:2009} \textrm{S. Fiori and T. Tanaka}, \textit{An algorithm to compute averages on matrix Lie groups}, IEEE Transactions on Signal Processing, Vol. 57, No. 12, pp. 4734 -- 4743, December 2009
%
%
\bibitem{flash} \textrm{T. Flash and N. Hogans}, \textit{The coordination of arm movements: An experimentally confirmed mathematical model}, The Journal of Neuroscience, Vol. 5, pp. 1688 -- 1703, 1985
%
\bibitem{Greenwood} \textrm{D.T. Greenwood}, \textit{Principles of Dynamics}, Second Edition, Prentice Hall, Upper Saddle River, NJ, 1988
%
\bibitem{he} \textrm{H. He, R. Li and K. Chen}, \textit{Characteristics of jerk response spectra for elastic and inelastic systems}, Shock and Vibration,
Vol. 2015, Article ID 782748, 12 pages, 2015
%
\bibitem{kerr} \textrm{A. Kerr, V.P. Pomeroy, P.J. Rowe, P. Dall and D. Rafferty}, \textit{Measuring movement fluency during the sit-to-walk task}, Gait \& Posture, Vol. 37, No. 4, pp. 598 -- 602, April 2013
%
\bibitem{kettunen} \textrm{J.A. Kettunen, U.M. Kujala, H. R\"{a}ty, T. Videman, S. Sarna, O. Impivaara and S. Koskinen}, \textit{Factors associated with hip joint rotation in former elite athletes}, British Journal of Sports Medicine, Vol. 34, pp. 44 -- 48, 2000
%
\bibitem{purdue} \textrm{A. Kim, J. Kim, S. Rietdyk and B. Ziaie}, \textit{Field assessment of gait: Accurate measures of step length and step length variability provided with a simple, inexpensive device}, \url{http://www.purdue.edu}, June 2014
%
\bibitem{mellone} \textrm{S. Mellone, C. Tacconi and L. Chiari}, \textit{Validity of a smartphone-based instrumented timed up and go}, Gait \& Posture, Vol. 36, No. 1, pp. 163 -- 165, May 2012
%
\bibitem{nishiguchi} \textrm{S. Nishiguchi, M. Yamada, K. Nagai, S. Mori, Y. Kajiwara, T. Sonoda, K. Yoshimura, H. Yoshitomi, H. Ito, K. Okamoto, T. Ito, S. Muto, T. Ishihara and T. Aoyama}, \textit{Reliability and validity of gait analysis by android-based smartphone}, Telemedicine and e-Health, Vol. 18, No. 4, pp. 292 -- 296, May 2012
%
\bibitem{postel}  \textrm{J. Postel}, \textit{RFC 768: User Datagram Protocol}, Internet Engineering Task Force at \verb+https://tools.ietf.org/html/rfc768+, August 1980
%
\bibitem{rubio} \textrm{F. Rubio, F. Valero, J. Sunyer and J. Cuadrado}, \textit{Optimal time trajectories for industrial robots with torque, power, jerk and energy consumed constraints}, Industrial Robot: An International Journal, Vol. 39, No. 1, pp. 92 -- 100, 2012
%
\bibitem{russell} \textrm{S.D. Russell, C.A. Zirker and S.S. Blemker}, \textit{Computer models offer new insights into the mechanics of rock climbing}, Sports Technology, Vol. 5, No. 3–4, pp. 120 -- 131, August-November 2012
%
\bibitem{seifert} \textrm{L. Seifert, D. Orth, J. Boulanger, V. Dovgalecs, R. H\'{e}rault and K. Davids}, \textit{Climbing skill and complexity of climbing wall design: Assessment of jerk as a novel indicator of performance fluency}, Journal of Applied Biomechanics, Vol. 30, pp. 619 -- 625, 2014
%
%
\bibitem{tack} \textrm{G.-R. Tack, J.S. Choi, J.H. Yi and C.H. Kim}, \textit{Relationship between jerk cost function and energy consumption during walking}, in Proceedings of the 2006 World Congress on Medical Physics and Biomedical Engineering, Vol. 14 of the series IFMBE Proceedings, pp. 2917 -- 2918, 2006
%
\bibitem{Tsao} \textrm{C.-C. Tsao and M.M. Mirbagheri}, \textit{Upper limb impairments associated with spasticity in neurological disorders}, Journal of NeuroEngineering and Rehabilitation, Vol. 4, No. 45, pp. 1 -- 15, 2007
%
\end{thebibliography}
\end{document}